\documentstyle[aps,prl,epsf]{revtex}

\begin{document}

\def\fr{\frac}   \def\dg{\dagger}   \def\b{\bar}  \def\s{\sum} \def\r{\rho}
\def\ga{\gamma}  \def\no{\nonumber}  \def\eq{\frac{e^4}{q^2}}
\def\lri{\longrightarrow} \def\si{\sigma} \def\cd{\cdot}
\def\de{\delta}
\def\th{\theta} \def\e{\epsilon}
\def\sq{sqrt}  \def\al{\alpha}  \def\Om{\Omega}
\def\sO{\fr{d\sigma}{d\Omega}} \def\sq{\sqrt} \def\up{\upsilon}
\def\inf{\infty} \def\la{\lambda}
\def\be{\beta}  \def\om{\omega} \def\De{\Delta}

\title{Bremsstrahlung Radiation as Coherent State \\
       in Thermal QED}

\author{Enke Wang, Jun Xiao and Hanzhong Zhang}

\address{Institute of Particle Physics, Huazhong Normal University,
         Wuhan 430079, China}

\date{\today}

\maketitle

\begin{abstract}
Based on fully finite temperature field theory we investigate the
radiation probability in the bremsstrahlung process in thermal
QED. It turns out that the infrared divergences resulting from the
emission and absorption of the real photons are canceled by the
virtual photon exchange processes at finite temperature. The full
quantum calculation results for soft photons radiation coincide
completely with that obtained in the semi-classical approximation.
In the framework of Thermofield Dynamics it is shown that the
bremsstrahlung radiation in thermal QED is a coherent state, the
quasiclassical behavior of the coherent state leads to above
coincidence.
\end{abstract}



\section{Introduction}
\label{sec1}

    An important aspect of the research on the quark gluon plasma (QGP)
is the evidence for its formation in ultrarelativistic heavy ion
collision. A lot of fruitful theoretical investigations have been
made on the signals for the existence of the QGP, for example,
$J/\Psi$ suppression \cite{Matsui}, strangeness
enhancement\cite{Rafelski}, jet quenching\cite{Wang1}, dilepton
and thermal photon production\cite{Ruuskanen} etc. As the
electromagnetic signal of the QGP the production of the thermal
photon has been investigated by calculating the Compton scattering
and $q {\bar q}$-annihilation processes in the QGP
state\cite{Kapusta,Baier,Arbex,Sollfrank,Cleymans}. Recent
research\cite{Aurenche} shows a very interesting result: the
bremsstrahlung processes in the QGP state play an important role
in the thermal photon production, it generates contribution of the
same order of magnitude as Compton scattering and $q {\bar
q}$-annihilation processes based on the resummation of hard
thermal loops\cite{Pisarski1}. So that the investigation of the
bremsstrahlung processes will be a subject of intense theoretical
interest. The investigation in Ref.\,\cite{Aurenche} focus on the
bresstralung processes with one photon radiation induced by the
parton-parton interaction inside QGP. In this paper We consider an
external charged particle prepared at remote past goes through the
plasma in thermal QED, which produces bresstralung radiation with
$N (\ge 1)$ photon radiation by colliding with charged constituent
inside the plasma. Such study will be of help to calculate the
energy loss of fermion when it goes through the
plasma\cite{Wang2}.

In the semi-classical approximation (classical charged current
coupled with a quantized electromagnetic field) Weldon
investigated the bremsstrahlung processes in QED and emphasized on
the cancellation of the infrared divergence\cite{Weldon}. In this
paper, however, we use fully finite temperature quantum field
theoretical method to analyze above bremsstrahlung processes by
calculating Feynman diagrams. we will show that the same
cancellation of the infrared divergence as in Ref.\, \cite{Weldon}
occurs in the quantum field theoretical calculation. An
interesting result is that quantum calculation results coincide
completely with the result obtained in the semi-classical
approximation for the radiation probability. We will argue the
reason why the quantum effects of the bremstrahlung can be ignored
by analyzing the connection between the electromagnetic field in
the bremsstrahlung process and coherent state at finite
temperature.

This paper is organized as follows: In Sec.~\ref{sec2} we derive
the radiation probability of the bremsstrahlung process by
calculating the Feynman diagram in the finite temperature field
theory and investigate the cancellation of the infrared
divergence. In Sec.~\ref{sec3} we demonstrate that at finite
temperature the radiation of soft photons in the bremsstrahlung
processes is a coherent state in the framework of Thermofield
Dynamics. A short summary is given in Sec.~\ref{sec4}.

\section{Radiation probability and infrared divergence cancellation}
\label{sec2}

Consider an electron that passes through a plasma in equilibrium
state with temperature $T$. The electron accelerates by colliding
with the charged constituent inside the plasma and produces the
bremsstrahlung radiation.

Now we consider the outgoing electron line in which the electron
radiates $n$ photons with momentum $k_1, k_2, \cdots, k_n$,
respectively. Denote the initial four-momentum of the electron as
$p$ and the final four-momentum as $p'$.For the moment we don't
care whether these are external real photons or virtual photons
which connect to each other or the vertices on the incoming
electron line. In the calculation we should sum over all possible
Feynman diagrams which correspond to $n!$ permutations for the
ordering of the momenta $k_1, k_2, \cdots, k_n$ as illustrated in
Fig.~1. If we denote such permutation as $\pi = (\pi(1), \pi(2),
\cdots, \pi(n) )$ and $\pi(i)$ is the number among $1$ and $n$,
then the Dirac structure (the part related to $\gamma$ matrices in
the invariant scattering amplitude) of the Feynman diagrams is
\begin{eqnarray}
\label{dirac1}
 \s\limits_{\{\pi\}} && \b u(p')(-ie\ga^{\mu_1})
    \fr{i({\not p}' +{\not k}_{\pi(1)}+m)}{(p'+k_{\pi(1)})^2-m^2}
    (-ie\ga^{\mu_2})
 \nonumber\\
   &&\fr{i({\not p}' +{\not k}_{\pi(1)}+{\not k}_{\pi(2)}+m)}
    {(p'+k_{\pi(1)}+k_{\pi(2)})^2-m^2}\cdots(-ie\ga^{\mu_n})
 \nonumber\\
   &&\fr{i({\not p}'+{\not k}_{\pi(1)}+\cdots+{\not k}_{\pi(n)}+m)}
   {(p'+k_{\pi(1)}+\cdots+k_{\pi(n)})^2-m^2}(i{\cal M}_{hard})\, ,
\end{eqnarray}
where $\{\pi\}$ denotes all possible permutation $\pi$. In
the following we consider only soft photons radiation. ${\cal
M}_{hard}$ denotes the invariant scattering amplitude of hard
processes in which no infrared divergence appears. For soft photon
emission, $k_i$ is small, we will drop the ${\not k}_i$ terms in
the numerators and the ${\cal O}(k^2)$ terms in the denominators.
It is shown\cite{Peskin} the Dirac structure (\ref{dirac1}) become
\begin{equation}
\label{dirac11}
   \fr{ep'^{\mu_1}}{p'\cd k_1}\fr{ep'^{\mu_2}}{p'\cd k_2}\cdots
   \fr{ep'^{\mu_n}}{p'\cd k_n}\b u(p') (i{\cal M}_{hard})\, .
\end{equation}

Similar to the above derivation, for $n$ photons attache to the
incoming electron line as illustrated in Fig.~2, the Dirac
structure of Feynman diagrams in Fig.~2 is deduced as
\begin{equation}
\label{dirac2}
   i{\cal M}_{hard}u(p)\fr{ep^{\mu_1}}{-p\cd k_1}\fr{ep^{\mu_2}}{-p\cd k_2}
   \cdots \fr{ep^{\mu_n}}{-p\cd k_n}\, .
\end{equation}

In general case, for $n$ soft photons emission if we don't
distinguish which one comes from the outgoing or incoming electron
line, the Dirac structure corresponding to the sum over all
possible Feynman diagrams can be expressed as
\begin{eqnarray}
\label{dirac3}
    \b u(p')& &i{\cal M}_{hard}u(p)\cd e(\fr{p'^{\mu_1}}{p'\cd k_1}
    -\fr{p^{\mu_1}}{p\cd k_1})
  \nonumber\\
    & &e(\fr{p'^{\mu_2}}{p'\cd k_2}-\fr{p^{\mu_2}}{p\cd k_2})
    \cdots e(\fr{p'^{\mu_n}}{p'\cd k_n}-\fr{p^{\mu_n}}{p\cd k_n})\, .
\end{eqnarray}

When calculating the radiation probability of the real photon
emission in the thermal equilibrium environment of the plasma, the
Boson enhancement factor $1+N(|{\bf k}|)$ should be taken into
account in the phase space integration. Here $N(|{\bf
k}|)=1/[\exp(|{\bf k}|)-1]$ is the Bose-Einstein distribution
function. For the emission of a photon with energy $\omega=|{\bf
k}|$, from Eq.~(\ref{dirac3}) we get the radiation probability
\begin{eqnarray}
\label{oneemi}
   \fr{dP_1}{d\om} & = & \int\fr{d\Om |{\bf k}|^2}{(2\pi)^3}
   \fr{e^2}{2|{\bf k}|}[1+N(|{\bf k}|)]
 \nonumber\\
   & &\cdot\sum\limits_{\la}(\fr{p'^\mu}{p'\cd k}-\fr{p^\mu}{p\cd k})
  (\fr{p'^\nu}{p'\cd k}-\fr{p^\nu}{p\cd k})\e_\mu^*(\la)\e_\nu(\la)
 \nonumber\\
   & = & \fr{1}{\om}[1+N(\om)]A(p',p)\, ,
\end{eqnarray}
where $\e_\mu(\la)$ is the polarization vector of the
photon and function $A(p',p)$ has been introduced in
Ref.\,\cite{Weldon},
\begin{eqnarray}
\label{A}
   A(p',p)&=&-\fr{e^2}{2}\int\fr{d\Om}{(2\pi)^3}
            (\fr{p'}{p'\cd\hat{k}}-\fr{p}{p\cd\hat{k}})
            \cd(\fr{p'}{p'\cd\hat{k}}-\fr{p}{p\cd\hat{k}})
 \nonumber\\
   &=&\fr{\al}{\pi}[\fr{1}{\up}\ln(\fr{1+\up}{1-\up})-2]\, ,
\end{eqnarray}
where $\hat{k}=(1, {\bf k}/|{\bf k}|)$ and $v$ is the
relative velocity of the final electron in the rest frame of the
initial electron defined by $p\cd p'=m^2/\sqrt{1-v^2}$.

>From Eq.~(\ref{oneemi}) we see clearly that the radiation probability of
a photon can be divided into two parts, the first part is the same as
obtained in quantum field theory and it is logarithmic infrared
divergent\cite{Peskin}. The second part is the contribution of
the thermal effects. As $\omega\rightarrow 0$, $N(\omega)\sim T/\omega$,
then the second part
\begin{equation}
\label{secpart}
 \fr{N(\om)}{\om}A(p',p) \sim \fr{T}{\om^2}A(p',p)\, ,
\end{equation}
which indicates a new infrared divergence emerged in finite temperature
field theory
and this infrared divergence is linear.

It is different to the quantum field theory at zero temperature, at finite
temperature the electron can absorb thermal photon from the heat bath. In
getting the absorption probability, the phase space integration should be
multiplied by a Boson absorption factor $N(|{\bf k}|)$. It is similar to
the above derivation, the probability for absorbing a photon with energy
$\omega=|{\bf k}|$ can be deduced as
\begin{equation}
\label{oneabs}
  \fr{dP'_1}{d\om}=\fr{1}{\om}N(\om)A(p',p)\, .
\end{equation}
We see again the linear infrared divergence resulting from the thermal
effects as $\om\rightarrow 0$.

As well known, in the QED theory at zero temperature the infrared
divergences to all order are canceled by taking into account all
possible virtual processes\cite{Bloch,Yennie,Weinberg}. In the
following we investigate whether the new infrared divergences
emerged in finite temperature field theory can be canceled by the
virtual processes.

By picking two photon momenta $k_i$ and $k_j$ and setting
$k_i=-k_j\equiv k$, we can make a virtual photon. At finite
temperature the convenient way for calculating the contribution of
virtual photon to the invariant scattering amplitude is to use the
real time formalism\cite{Bellac}, for single virtual photon
exchange process we have
\begin{equation}
\label{X1}
    X={1\over 2}\int\frac{d^4 k}{(2\pi)^4}{e^2}
     (\frac{p'^{\mu_i}}{p'\cd k}-\frac{p^{\mu_i}}{p\cd k})
     (\frac{p'^{\mu_j}}{-p'\cd k}-\frac{p^{\mu_j}}{-p\cd k})
     \Delta^{11}_{\mu_i\mu_j}(k)\, ,
\end{equation}
where $1/2$ is needed because our procedure counted each
Feynman diagram twice, $\Delta^{11}_{\mu_i\mu_j}(k)$ is the
(11)-component of the photon matrix propagator in the real time
formalism,
\begin{equation}
\label{Delta}
    \Delta^{11}_{\mu_i\mu_j}(k)=-g_{\mu_i\mu_j}
    \left( {i\over {k^2+i\eta}}+2\pi N(|k_0|)\delta(k^2)
    \right)\, .
\end{equation}
The real part of $X$ arises from
$(1+2N(|k_0|))\pi\delta(k^2)$ term, so that we get
\begin{eqnarray}
\label{X}
    \Re(X)&=&{e^2\over 2}\int\frac{d^4 k}{(2\pi)^4}
     (\frac{p'}{p'\cd k}-\frac{p}{p\cd k})\cdot
     (\frac{p'}{p'\cd k}-\frac{p}{p\cd k})
     (1+2N(|k_0|))\pi\delta(k^2)
\nonumber\\
    &=&-\fr{1}{2}A(p',p)\int\limits
        \fr{d|{\bf k}|}{|{\bf k}|}\bigl(1+2N(|{\bf k}|)\bigr)\, .
\end{eqnarray}
Similar to Eq.~(\ref{oneemi}), $\Re(X)$ can be divided into
two parts, one part is the same as obtained in quantum field
theory and another part is the contribution of the thermal
effects. Both parts are all infrared divergent and give negative
contribution comparing to the real emission~(\ref{oneemi}) and
absorption~(\ref{oneabs}).

The effect of adding one virtual infrared-photon correction to a
diagram which does not involve any virtual processes is to
multiply the amplitude by a factor $(1+X)$, and thus multiplies
the rate by $|1+X|^2\approx 1+2\Re(X)$. For $n$ virtual photon
exchange processes, each virtual photon will contribute a $X$ to
the invariant scattering amplitude, sum over $n$, then the total
contribution of all possible virtual photons to the radiation
probability can be exponentiated as
\begin{equation}
 \left| \sum \limits_{n=0}^{\infty}\fr{X^n}{n!}\right|^2
   = e^{2\Re(X)}\, ,
\end{equation}
where $1/n!$ is a symmetry factor because interchanging virtual photon with
each other does not change the diagram.

In the practical calculation of the bremsstrahlung process, all possible
emissions and absorptions of the real photons and all possible exchanges
of the virtual photon should be taken into account at the same time. So that
the probability of radiating a net energy $E$ is
\begin{eqnarray}
\label{radposs}
     \fr{dP}{dE}&=&\exp[2\Re(X)]\sum\limits_{n=1}^{\infty}\int\fr{1}{n!}
     d\Phi_1 d\Phi_2 \cdots d\Phi_n
  \nonumber\\
    &&\de(k^0_1+k^0_2+\cdots +k^0_n-E)
      {\prod\limits_{i=1}^{n}}\sum\limits_{\la}
      e^2 (\fr{p'^\mu}{p'\cd k_i}-\fr{p^\mu}{p\cd k_i})
  \nonumber\\
    &&(\fr{p'^\nu}{p'\cd k_i}-\fr{p^\nu}{p\cd k_i})\e_\mu^*(\la)\e_\nu(\la)\, ,
\end{eqnarray}
where $1/n!$ is symmetry factor because $n$ photons are identical particles,
and $d\Phi_i$ is integration of phase space. For emission
($k^0_i=|{\bf k}_i|$) and absorption ($k^0_i=-|{\bf k}_i|$), $d\Phi_i$ is
defined as
\begin{eqnarray}
\label{phase}
    d\Phi_i &=& \fr{d^3|{\bf k}_i|}{(2\pi)^3}\fr{1}{2|{\bf k}_i|}\times
              \cases{
               1+N(|{\bf k}_i|),  & if $ k^0_i=|{\bf k}_i|$ \cr
               N(|{\bf k}_i|),    & if $ k^0_i=-|{\bf k}_i|$ \cr}
  \nonumber\\
   & = & \fr{d^4 k_i}{(2\pi)^3}\de(k_i^2)[\th(k_i^0)+N(|{\bf k}_i|)]\, .
\end{eqnarray}
>From Eq.~(\ref{radposs}) it is easy to deduce
\begin{equation}
\label{radposs1}
   \fr{dP}{dE}=
   \int\limits_{-\infty}^{+\infty}\fr{dt}{2\pi}e^{-iE t}e^{\b R(t)}\, ,
\end{equation}
where
\begin{eqnarray}
   \b R(t) &=& 2\Re(X)+R(t)\, ,
  \label{R1}\\
    R(t) &=& -e^2\int d\Phi_i e^{ik^0_i t}(\fr{p'}{p'\cd k_i}-\fr{p}{p\cd k_i})
  \nonumber\\
    &&\qquad\cd(\fr{p'}{p'\cd k_i}-\fr{p}{p\cd k_i})\, .
  \label{R2}
\end{eqnarray}
Both $\Re(X)$ and $R(t)$ contain infrared divergence, but
the sign is opposite. Inserting Eqs.~(\ref{X}) and (\ref{R2}) into
Eq.(\ref{R1}) we get
\begin{eqnarray}
\label{R3}
   {\b R(t)} & = & A(p', p)\int
     \fr{d |{\bf k}|}{|{\bf k}|}\Bigl[\bigl(1+N(|{\bf k}|)\bigr)
     e^{i|{\bf k}|t}
 \nonumber\\
    &&\qquad +N(|{\bf k}|)e^{-i|{\bf k}|t}
     -\bigl(1+2N(|{\bf k}|)\bigr)\Bigr]\, .
\end{eqnarray}
If $|{\bf k}|$ is small, then $ N(|{\bf k}|)\sim T/|{\bf
k}|$ and $e^{i|{\bf k}|t}\sim 1+i|{\bf k}|t$. Substitute them into
Eq.~(\ref{R3}), the linear and the logarithmic divergent terms are
canceled exactly, so that the radiation probability
(\ref{radposs1}) is finite in the infrared region at finite
temperature.

In order to perform the integration in Eq.~(\ref{R3}) it is convenient to
introduce a cutoff factor $\exp(-|{\bf k}|/\Lambda)$ to the integrand
in the ultraviolet region. After introducing this cutoff factor,
the radiation probability (\ref{radposs1}) coincides completely with
the result,
\begin{equation}
\label{dpde}
    \fr{dP}{dE}=\left|\Gamma\left(\fr{A}{2}+i\fr{E}{2\pi T}\right)\right|^2
               \fr{e^{E/2T}e^{-|E|/\Lambda}}{4\pi^2 T\Gamma(A)}
               \left(\fr{2\pi T}{\Lambda}\right)^A\, ,
\end{equation}
which has been obtained in Ref.~\cite{Weldon} in the semi-classical
approximation.

\section{Connection between bremsstrahlung radiation and coherent state}
\label{sec3}

In above full quantum field theoretical calculation we consider only
soft photons. The full quantum calculation result is
completely the same as that obtained in the semi-classical approximation.
It seems that
the quantum effects is unimportant for soft photon radiation in
bremsstrahlung process. In the following we argue the reason.

Now we calculate the probability of
producing $n$ photons in bremsstrahlung
process. Suppose the energy of all photons is between $\om$ and
$\om +\Delta\om$. The contribution of all virtual photons to
probability amplitude is
\begin{equation}
\label{virtual}
     \Re(X')=-\fr{1}{2}A(p',p)\int\limits_{\om}^{\om+\rm\Delta\om}
        \fr{d|{\bf k}|}{|{\bf k}|}\bigl(1+2N(|{\bf k}|)\bigr)\, .
\end{equation}
The probability for producing $n$ photon with energy between $\om$ and
$\om +\Delta\om$ can be expressed as
\begin{eqnarray}
\label{pn}
   P_n(\om) &=&\exp[2\Re(X')]\fr{1}{n!}{\prod\limits_{i=1}^n}\sum_{\la}
     \int\limits_{\om}^{\om+\rm \Delta\om}d\Phi_i
 \nonumber\\
    & &e^2(\fr{p'^{\mu}}{p'\cd k_i}-\fr{p^{\mu}}{p\cd k_i})
    (\fr{p'^{\nu}}{p'\cd k_i}-\fr{p^{\nu}}{p\cd k_i})
     \e_\mu^*(\la)\e_\nu(\la)
 \nonumber\\
    &=&{{\la^n}\over n!}e^{-\la}\, .
\end{eqnarray}
It is a Poisson distribution. The average photon number is
\begin{equation}
\label{n}
    \langle n\rangle=\la=A(p',p)\int\limits_{\om}^{\om+\rm \Delta\om}
    \fr{d |{\bf k}|}{|{\bf k}|}[1+2N(|{\bf k}|)]\, .
\end{equation}
At high temperature limit $T\gg 1$, $N(|{\bf k}|)\sim T/|{\bf
k}|$, we see that the average photon number increases linearly
with increasing temperature.

It is known that at zero temperature the final state in the
bremsstrahlung process is a coherent state and the distribution of
the photon are also a Poisson distribution\cite{Itzykson} like
Eq.(\ref{pn}). This stimulate us to investigate whether the
bremsstrahlung radiation result from a coherent state at finite
temperature.

In the following we demonstrate that at finite temperature the
photon radiation in the bremsstrahlung process really does come
from a coherent state of the field operator in the framework of
Thermofield Dynamics\cite{Umezawa}. In the Thermofield Dynamics a
so-called tilde system which is identical to the original system
has been introduced in order to define the thermal vacuum state
and the thermal field operator. The complete Lagrangian for the
combined system describing the radiation of photons by a classical
current $j_{\mu}(x)$ is
\begin{equation}
\label{L}
    {\hat {\cal L}}={\cal L}-{\tilde {\cal L}}\, ,
\end{equation}
where the Lagrangian for original and tilde system are
\begin{eqnarray}
\label{L1}
   {\cal L}&=&-\left({1\over 4}
   F_{\mu\nu}F^{\mu\nu}+j_{\mu}A^{\mu}\right)\, ,
\\
   {\tilde {\cal L}}&=&-\left({1\over 4}
   {\tilde F}_{\mu\nu}{\tilde F}^{\mu\nu}
   +{\tilde j}_{\mu}{\tilde A}^{\mu}\right)\, ,
\end{eqnarray}
respectively. In Thermofield Dynamics we introduce the doublet
field and source
\begin{equation}
\label{AJ}
   {\cal A}_{\mu}(x) = \left( \begin{array}{cc}
                        A_{\mu}(x)\\
                        {\tilde A}_{\mu}(x)
                       \end{array} \right)\, , \qquad
   {\cal J}_{\mu}(x) = \left( \begin{array}{cc}
                         j_{\mu}(x)\\
                         {\tilde j}_{\mu}(x)
                        \end{array} \right)\, .
\end{equation}
For the bremsstrahlung process we can assume that the current
${\cal J}_{\mu}(x)$ is switched on adiabatically on a finite time
interval. Use the retarded and advanced Green function the
solution of the equation of motion for field ${\cal A}_{\mu}(x)$
at zero temperature can be expressed as
\begin{eqnarray}
\label{A1}
   {\cal A}_{\mu}(x) &=& {\cal A}_{\mu}^{in}(x)+
     \int d^4 y G_{\mu\nu}^{ret}(x-y){\cal J}^{\nu}(y)
\nonumber\\
     &=& {\cal A}_{\mu}^{out}(x)+
     \int d^4 y G_{\mu\nu}^{adv}(x-y){\cal J}^{\nu}(y)\, .
\end{eqnarray}
The quantum free field ${\cal A}_{\mu}^{in}(x)$ and ${\cal
A}_{\mu}^{out}(x)$ describe the photon field before and after its
interaction with the current ${\cal J}_{\mu}$, then we have
\begin{equation}
\label{Ainout}
   {\cal A}_{\mu}^{in}(x)=\lim_{x_0\rightarrow -\infty}
   {\cal A}_{\mu}(x)\, ,
   \qquad
   {\cal A}_{\mu}^{out}(x)=\lim_{x_0\rightarrow +\infty}
   {\cal A}_{\mu}(x)\, .
\end{equation}
At zero temperature the retarded and advanced Green function can
be written as
\begin{eqnarray}
\label{Gretadv}
   i G_{\mu\nu}^{ret}(x-y) &=& \theta(t_x-t_y)\langle 0,{\tilde 0}|
     \left( \begin{array}{cc}
        [A_{\mu}^{(0)}(x), A_{\nu}^{(0)}(y)] , & 0\\
        0, & [{\tilde A}_{\mu}^{(0)}(x), {\tilde A}_{\nu}^{(0)}(y)]
     \end{array} \right)|0,{\tilde 0}\rangle\, ,
\\
   i G_{\mu\nu}^{adv}(x-y) &=& -\theta(t_y-t_x)\langle 0,{\tilde 0}|
     \left( \begin{array}{cc}
        [A_{\mu}^{(0)}(x), A_{\nu}^{(0)}(y)] , & 0\\
        0, & [{\tilde A}_{\mu}^{(0)}(x), {\tilde A}_{\nu}^{(0)}(y)]
     \end{array} \right)|0,{\tilde 0}\rangle\, .
\end{eqnarray}

In the Thermofield Dynamics the connection between usual vacuum
$|0,{\tilde 0}\rangle$ and the thermal vacuum $|0,\beta\rangle$ is
a Bogoliubov transformation\cite{Umezawa},
\begin{eqnarray}
\label{thermalvac}
   |0,\beta\rangle&=&U(\theta)|0,{\tilde 0}\rangle\, ,
\\
   U(\theta)&=&\exp\left[-\sum_{\bf k}\theta_{\bf k}(\beta)
     \left({\tilde a}_{\bf k}a_{\bf k}-a^{\dag}_{\bf k}
     {\tilde a}_{\bf k}^{\dag}\right)\right]\, ,
\end{eqnarray}
with
\begin{equation}
\label{theta}
   \cosh\theta_{\bf k}(\beta)=
    {1\over \sqrt{1-e^{-\beta |{\bf k}|}}}\, , \qquad
   \sinh\theta_{\bf k}(\beta)=
    {e^{\beta|{\bf k}/2}\over \sqrt{1-e^{-\beta |{\bf k}|}}}\, ,
\end{equation}
where $\beta=1/T$. From Eq.(\ref{A1}) corresponding thermal field
can be expressed as
\begin{eqnarray}
\label{A2}
   {\cal A}_{\mu}(x,\beta)&=&U^{\dag}(\theta)
     {\cal A}_{\mu}(x)U(\theta)
\nonumber\\
   &=& {\cal A}_{\mu}^{in}(x,\beta)+
     \int d^4 y G_{\mu\nu}^{ret}(x-y,\beta)
     {\cal J}^{\nu}(y,\beta)
\nonumber\\
     &=& {\cal A}_{\mu}^{out}(x,\beta)+
     \int d^4 y G_{\mu\nu}^{adv}(x-y,\beta)
     {\cal J}^{\nu}(y,\beta)\, ,
\end{eqnarray}
where corresponding thermal field, retarded and advanced Green
function and the source are
\begin{eqnarray}
\label{thermalf}
   {\cal A}_{\mu}^{in,out}(x,\beta)&=&U^{\dag}(\theta)
     {\cal A}_{\mu}^{in,out}(x)U(\theta)\, ,
\\
   G_{\mu\nu}^{ret,adv}(x-y,\beta)&=&U^{\dag}(\theta)
     G_{\mu\nu}^{ret,adv}(x-y)U(\theta)\, ,
\\
   {\cal J}_{\mu}(x,\beta)&=&U^{\dag}(\theta)
     {\cal J}_{\mu}(x)U(\theta)\, .
\end{eqnarray}
>From Eq.(\ref{A2}) we obtain
\begin{eqnarray}
\label{A3}
   {\cal A}_{\mu}^{out}(x,\beta)&=&
     {\cal A}_{\mu}^{in}(x,\beta)+
     \int d^4 y \left[G_{\mu\nu}^{ret}(x-y,\beta)-
     G_{\mu\nu}^{adv}(x-y,\beta)\right]
     {\cal J}^{\nu}(y,\beta)
\nonumber\\
     &=&{\cal A}_{\mu}^{in}(x,\beta)
      +{\cal A}_{\mu}^{cl}(x,\beta)\, ,
\end{eqnarray}
where ${\cal A}_{\mu}^{cl}(x,\beta)$ is the classical thermal
field radiated by the current ${\cal J}(x,\beta)$ at finite
temperature.

Define the thermal in- and out-state connecting the thermal in-
and out-state field as $|0, \beta\rangle_{in}$ and $|0,
\beta\rangle_{out}$, respectively. They are connected by a unitary
operator $S$ as
\begin{equation}
\label{inout}
   |0, \beta\rangle_{out}=S |0, \beta\rangle_{in}\, .
\end{equation}
Then we have
\begin{equation}
\label{inout2}
   {\cal A}_{\mu}^{out}(x,\beta)=
   S^{-1}{\cal A}_{\mu}^{in}(x,\beta)S\, .
\end{equation}
The thermal vacuum state $|0, \beta\rangle_{in}$ is an eigenvector
of the annihilation part ${\cal A}_{\mu}^{in(+)}(x,\beta)$ with
positive frequency,
\begin{equation}
\label{in}
   {\cal A}_{\mu}^{in(+)}(x,\beta)|0, \beta\rangle_{in}=0\, .
\end{equation}
>From Eq.(\ref{A3}) we have
\begin{eqnarray}
\label{out}
   {\cal A}_{\mu}^{out(+)}(x,\beta)|0, \beta\rangle_{in}
   &=&S^{-1}{\cal A}_{\mu}^{in(+)}(x,\beta)S|0, \beta\rangle_{in}
\nonumber\\
   &=&{\cal A}_{\mu}^{cl(+)}(x,\beta)|0, \beta\rangle_{in}\, .
\end{eqnarray}
This expression lead to
\begin{equation}
\label{coh}
   {\cal A}_{\mu}^{in(+)}(x,\beta)|0, \beta\rangle_{out}=
   {\cal A}_{\mu}^{cl(+)}(x,\beta)|0, \beta\rangle_{out}\, .
\end{equation}
Eq.(\ref{coh}) indicate that at finite temperature the final state
in bremsstrahlung process is a coherent state because it is just
the eigenstate of annihilation thermal field operator.

The main features of the coherent state are that the coherent
state is represented by a minimum uncertainty wave packet, the
quantum correlation in these states is absent, so that the
coherent state behaves as a quasiclassical state. It is the
quasiclassical property of the coherent state which leads to that
our full quantum calculation results coincides completely with
that obtained in the semi-classical approximation.

\section{Conclusion}
\label{sec4}

By directly calculating the Feynman diagram, we use fully finite
temperature field theoretical method to derive the radiation
probability of the bremsstrahlung processes in thermal QED. It is
shown that the infrared divergence are canceled to each other by
taking into account the real photon emission, absorption and
virtual photon exchange processes at the same time. The radiation
probability for soft photons coincides completely with the result
obtained in the semi-classical approximation. At zero temperature
it is known that the bremsstrahlung radiation comes from the
coherent state, we generalize this argument to the finite
temperature case. Based on the Thermofield Dynamics it is shown
that the final state in bremsstrahlung process is a coherent state
at finite temperature. Because the coherent state is a state which
is the most approachable to classical limit and is permitted to
exist in quantum theory, this quasiclassical features makes that
the semi-classical approximation produces same result as the
quantum calculation.

\acknowledgments
This work was supported in part by the National Natural Science Foundation
of China (NSFC) under Grant No. 19945001 and 19928511 and the Science Research
Foundation of Hubei Province in China.


 \begin{figure}
 \vskip 2cm
 \epsfxsize 150mm \epsfbox{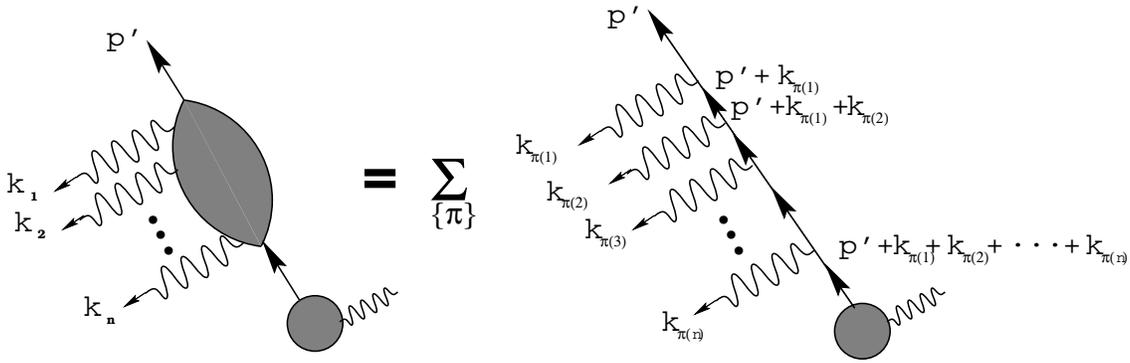}
 \vskip 0.4cm
 \caption{
   Feynman diagram for the outgoing electron line radiating $n$ photons.
 \label{F1}}
 \end{figure}

 \begin{figure}
 \vskip 2cm
  \epsfxsize 150mm \epsfbox{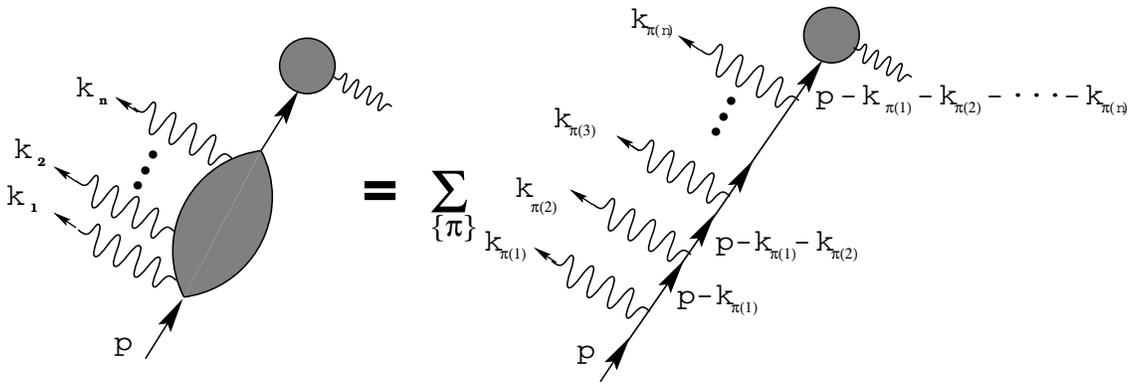}
  \vskip 0.4cm
  \caption{
   Same as Fig.~1 but for incoming electron line.
  \label{F2}}
 \end{figure}

\end{document}